\documentclass[aps,pra,amsmath,amssymb,reprint]{revtex4-2}

\usepackage{graphicx}
\usepackage{dcolumn}
\usepackage{bm}
\usepackage[referable]{threeparttablex}
\usepackage{array,booktabs}
\usepackage{multirow}

\usepackage{color}
\usepackage[dvipsnames]{xcolor}
\usepackage{caption}
\usepackage{subcaption}

\DeclareMathAlphabet{\mathcal}{OMS}{cmsy}{m}{n}
\usepackage[breaklinks=true,colorlinks=true,linkcolor=blue,urlcolor=blue,citecolor=blue,bookmarks=false]{hyperref}
\begin{document}

\title[]{Parity-violating nuclear spin-rotation and NMR shielding tensors in tetrahedral molecules}

\author{I.~Agust{\'{\i}}n Aucar}
\email[Author to whom correspondence should be addresed. Electronic mail: ]{agustin.aucar@conicet.gov.ar}
\affiliation{Instituto de Modelado e Innovaci\'on Tecnol\'ogica (UNNE-CONICET), Facultad de Ciencias Exactas y Naturales y Agrimensura, Universidad Nacional del Nordeste, Avda.~Libertad 5460, Corrientes, Argentina}

\author{Yuly Chamorro}
\author{Anastasia Borschevsky}
\affiliation{Faculty of Science and Engineering, Van Swinderen Institute for Particle Physics and Gravity, University of Groningen, 9747 AG Groningen, The Netherlands}

\date{\today}

\begin{abstract}

In natural processes involving weak interactions, a violation of spatial parity conservation should appear. Its effects are expected to be observable in molecules using different spectroscopic methodologies, but due to the tiny magnitude of these effects they have never been measured yet. We present a theoretical analysis of four-component relativistic nuclear-spin-dependent parity-violating nuclear spin-rotation and NMR shielding tensors in a set of tetrahedral chiral molecules. This work highlights the crucial role played by the ligands and the  electronic structure of the chiral center in the enhancement of these effects, leading the way towards targeted design of promising molecules for measurements.

\end{abstract}

\maketitle

\section{Introduction}\label{sec:intro}

The parity non-conserving nature of physical phenomena affected by weak interactions is known to be responsible for asymmetric processes with respect to spatial inversion of the coordinates of the particles in a system. The existence of parity violation (PV) was first postulated by Lee and Yang\cite{LeeYang56}, and it was Wu~\textit{et~al.}\cite{Wu57}~who provided its pioneer experimental verification using $^{60}$Co nuclei and observing nuclear $\beta$-decay processes due to weak interactions. While the first confirmations of the existence of PV effects were carried out in nuclear systems\cite{Wu57,Garwin57}, since then these phenomena were also observed in atoms\cite{Conti79,Barkov80,Bouchiat82,Emmons83,Macpherson91,WooBenCho97,Tsigutkin2009,Bouchiat12}.

Parity violating effects due to the weak interactions between electrons and nucleons are expected to be observable even in molecular systems. In the case of the two enantiomers of a chiral molecule, for example, a tiny energy difference is expected to be found due to the PV effects. Many experiments, using different methods, have been proposed to observe molecular PV effects\cite{Letokhov75,Kompanets76,Szabo99,Berger04-bookchapter,MacDermott04,DeMille08,DarStoShe10,Hobi2013,Cahn14,CouManPie19}, but in spite of the ever-improving precision of such experimental searches none of them was successful so far. Several theoretical studies addressed the influence of PV effects on nuclear magnetic resonance (NMR) parameters like shielding or indirect spin-spin coupling tensors in molecular systems\cite{Gorshkov82,Barra1986,Barra1988,Barra1996,Soncini2003,Laubender2003,Weijo2005,LauBer06,Bast2006,Nahrwold2009}, and first such measurements were recently attempted\cite{Eills2017,Blanchard2020}.

In this work we focus our attention on a property measured in microwave spectroscopy, namely, the nuclear spin-rotation (NSR) tensor, where molecular PV effects could be observed by analyzing the difference between the measurements of this parameter in two enantiomers of a chiral molecule. When the nuclear-spin-dependent (NSD) PV Hamiltonian is considered, a difference between the NSR tensors $\bm{M}$ for the nuclei in the two enantiomers of a chiral molecule is expected to appear, but if the PV effects are neglected, the $\bm{M}$ values of the two enantiomers will be equal. The difference appearing as a PV effect is due to an asymmetry in the electronic environments around the nuclei, arising from weak interactions. As suggested by Barra and co-workers, PV effects will contribute not only to the NSR constants, but also to the NMR shielding and indirect spin-spin coupling tensors\cite{Barra1986}. An experimental observation of PV effects in one of these parameters would give rise to the first detection of PV interactions in a static system. Atomic experiments to detect weak interactions, in contrast, need transition phenomena\cite{Bouchiat1997}.

In this paper we explore the PV effects on the parity-violating nuclear spin-rotation (PV-NSR) tensors of the central nuclei of tetrahedral chiral molecules, focusing on the role of the electronic environment and the atomic number of the central metal atom. We use the relativistic theoretical formalism we have recently derived\cite{AucarBorschevsky} based on previous investigations on parity-conserving NSR tensors within the relativistic domain\cite{Agus2012,Agus2013-1,Agus2013-2}.

We analyze the $^{183}$W and $^{235}$U nuclei in the NW$XYZ$ and NU$XYZ$ series of molecules, respectively (with $X,Y,Z=$ $^1$H, $^{19}$F, $^{35}$Cl, $^{79}$Br, $^{127}$I), and also the $X$ nuclei in the N$X$HF$Z$ molecules ($X=$ $^{53}$Cr, $^{77}$Se, $^{95}$Mo, $^{125}$Te; $Z=$ $^{35}$Cl, $^{79}$Br, $^{127}$I). We also report values of four-component (4c) isotropic PV nuclear shielding constants. Some of the selected systems were used in previous works to study PV effects in shielding and in other molecular properties\cite{Figgen2010,Figgen2010ACIE,Wormit2014,Nahrwold2014}.

We present a systematic analysis of relativistic and electronic correlation effects on the PV-NSR and PV-shielding tensors, employing the polarization propagator theory to calculate the corresponding linear response functions, within the random phase approximation (RPA) theory based on the Dirac-Coulomb (DC) Hamiltonian. Dirac-Hartree-Fock (DHF) and Dirac-Kohn-Sham (DKS) methodologies were used to obtain the wave functions. In order to compare relativistic 4c calculations with their non-relativistic (NR) limit, we also employed the Lévy-Leblond (LL) Hamiltonian.

This work has the following structure: In Sec.~\ref{sec:theory} we give a brief theoretical background to introduce the relativistic formulation used to calculate the NSD-PV-NSR and NSD-PV-shielding tensors. In Sec.~\ref{sec:comp-det}, we provide the computational details for all the calculations presented in this paper, divided in two main parts: geometry optimizations and linear response calculations, including a short analysis of the basis set convergence of the PV-NSR constants in one of the studied molecules. In Sec.~\ref{sec:results}, we present the 4c computations of the isotropic NSD-PV-NSR constants for all the molecules studied here. The relativistic and electronic correlation effects are studied in this Section, as well as the influence of different ligands (keeping the same metal nucleus or chiral center) and different chiral centers for the same ligands. Sec.~\ref{sec:conclusion} contains the conclusions of this work and the outlook.

\section{Theory}\label{sec:theory}

Within the polarization propagator theory, any static (\textit{i.e.}, zero frequency) second-order molecular property can be calculated as\cite{Oddershede1978,Aucar2014}
\begin{equation}\label{eq:E2-PolProp}
E^{(2)}_{PQ}= \textnormal{Re}\left[  \langle\langle \, \hat{H}^P \, ; \, \hat{H}^Q \, \rangle\rangle_{\omega=0}  \right],
\end{equation}
\noindent where the operators $\hat{H}^P$ and $\hat{H}^Q$ are any perturbative Hamiltonians. The linear response function on the right-hand-side of Eq.~\eqref{eq:E2-PolProp} can also be written as the product of the perturbator $\textbf{b}^P$ (i.e., the property matrix element), the principal propagator $\textbf{M}^{-1}$ (i.e., the inverse of the electronic Hessian), and the perturbator $\textbf{b}^Q$\cite{Aucar2010,AucarBorschevsky}. The linear response function is usually computed by solving the response equation
\begin{equation}\label{eq:resp-eq}
\textbf{M} \; \textbf{X}^Q(\omega) = \textbf{b}^Q,
\end{equation}
\noindent where $\textbf{X}^Q(\omega) = \textbf{M}^{-1} \, \textbf{b}^Q$ is expanded in a linear combination of trial vectors, and then it is contracted with the property matrix $\textbf{b}^P$\cite{Saue2003}.

In the particular case of the nuclear spin-rotation and the NMR shielding tensors of a nucleus $N$ ($\bm{M}_N$ and $\bm{\sigma}_N$, respectively) it is known that they can be obtained as the second order energy derivatives at zero frequency\cite{Agus2012,aucar-aucar2019}
\begin{eqnarray}
    \bm{M}_N &=& - \hslash \frac{\partial^2 E(\bm{I}_N,\bm{J})}{\partial \bm{I}_N \partial \bm{J}} \bigg|_{\bm{I}_N=\bm{J}=0} \label{eq:M-energy} \\
    \bm{\sigma}_N &=& \frac{\partial^2 E(\bm{\mu}_N,\bm{B}_0)}{\partial \bm{\mu}_N \partial \bm{B}_0}\bigg|_{\bm{\mu}_N=\bm{B}_0=0} \label{eq:sigma-energy}
\end{eqnarray}
\noindent where $\hslash=\frac{h}{2\pi}$ is the reduced Planck constant, $\bm{I}_N$ is the dimensionless spin of nucleus $N$, $\bm{J}$ is the molecular rotational angular momentum around the molecular center of mass,  $\bm{B}_0$ is a uniform external magnetic field, and $\bm{\mu}_N = \gamma_N \, \hslash \, \bm{I}_N$ the magnetic moment due to the nuclear spin, where $\gamma_N = \frac{e}{2 m_p} g_N$ is the gyromagnetic ratio of nucleus $N$ and $g_N$ is its $g$-factor, and where $e$ is the fundamental charge and $m_p$ is the proton mass. The nuclear spin-rotation (NSR) tensor $\bm{M}_N$ in Eq.~\eqref{eq:M-energy} is given in units of energy. Therefore, by dividing it by the Planck constant $h$ the corresponding frequency values are obtained. SI units are used in the present work.

The PV contribution to the NSR tensor was recently derived within a relativistic framework, and it has been shown that it can be written as the linear response function\cite{AucarBorschevsky}
\begin{eqnarray}\label{eq:M-PV}
 \bm{M}_N^{PV} &=& \frac{\hslash \, G_F}{2\sqrt{2} \, c_0} \kappa_N 
 \langle\langle \; \rho_N(\bm{r}) \; c \bm{\alpha} \; ; \;  \bm{J}_e \; \rangle\rangle \; \cdot \bm{I}^{-1},
\end{eqnarray}
\noindent where $G_F$ is the Fermi coupling constant, whose most recent value is $G_F/(\hslash \, c_0)^3=1.1663787 \times 10^{-5}$~GeV$^{-2}$, or equivalently $G_F\simeq 2.222516 \times10^{-14} \, E_h \, a_0^3$\cite{codata2018}.

Besides, $\kappa_N = -2\lambda_N \left( 1 - 4 \, \textnormal{sin}^2 \theta_W \right)$, where $\lambda_N$ is a nuclear state dependent parameter. It can be noted that the constant factor in the NSD-PV Hamiltonian is found to be written in different ways in the literature\cite{Gorshkov1982,Barra1986,Barra1988,Nahrwold2009,Borschevsky2012,Isaev2012,Nahrwold2014,AucarBorschevsky}. There are three predominant contributions to $\lambda_N$ due to the following interactions\cite{Khriplovich1991,Ginges2004}: the weak coupling between neutral electronic vectors and nucleon axial-vector currents\cite{Novikov77}, the electromagnetic interactions between electrons and nuclear anapole moments (which become the dominant contributions for heavy nuclei)\cite{Flambaum1980,Flambaum84,Flambaum85-pla}, and the nuclear-spin-independent electron axial-vector and nucleon vector current weak interactions combined with hyperfine interactions\cite{Flambaum85}.

As $\lambda_N$ is a factor with a nuclear structure origin and to facilitate the comparison with previous works, we have set $\lambda_N=1$ in our calculations. Therefore, all the reported values of $\bm{M}_N^{PV}$ and $\bm{\sigma}_N^{PV}$ must be scaled by the true value of $\lambda_N$ in order to get results that can be compared with measurable physical quantities. For heavy nuclei it is expected that $1<\lambda_N<10$\cite{Flambaum1980,Flambaum84}.

While the most recent value of the sine-squared weak mixing angle $\theta_W$ is 0.23857(5)\cite{Zyla2020}, we use $\textnormal{sin}^2\theta_W=0.2319$\cite{sintheta} as the Weinberg parameter throughout this work for ease of comparison with earlier investigations\cite{AucarBorschevsky}.

In addition, in Eq.~\eqref{eq:M-PV} $\bm{\alpha}$ are the $4 \times 4$ Dirac matrices given in the standard representation based on the Pauli spin matrices, $\bm{r}$ is the position of the electrons with respect to the coordinate origin, $\rho_N(\bm{r})$ is the normalized nuclear electric charge density of nucleus $N$ at the position of electron (given in units of the inverse of cube distances), $\frac{1}{c_0}$ is linearly proportional to the fine structure constant (in SI units, the fine structure constant is $\frac{1}{4\pi \epsilon_0} \frac{e^2}{\hslash \, c_0}$), $c$ is the speed of light in vacuum, scalable to infinity at the NR limit, $\bm{I}^{-1}$ is the inverse molecular inertia tensor with respect to the molecular center of mass (CM) in the equilibrium geometry, and $\bm{J}_e = \bm{L}_e + \bm{S}_e$ is the $4 \times 4$ total electronic angular momentum operator.

In the present work we neglect the contributions to Eq.~\eqref{eq:M-PV} due to the Breit electron-nucleus interaction. These were shown to be very small for the parity-conserving NSR tensors\cite{Agus2013-2}.

Similarly to Eq.~\eqref{eq:M-PV}, the  PV contribution to the NMR shielding tensor is given by\cite{Barra1988,Bast2006}
\begin{eqnarray}\label{eq:sigma-PV}
 \bm{\sigma}_N^{PV} 
 &=& \frac{m_p\,G_F}{2\,\sqrt{2}\,\hslash\,c_0} \; \frac{\kappa_N}{g_N} \langle\langle \; \rho_N(\bm{r}) \; c \bm{\alpha} \; ; \;  \bm{r}_{GO} \times c \bm{\alpha} \; \rangle\rangle,
\end{eqnarray}
\noindent where $\bm{r}_{GO}=\bm{r}-\bm{R}_{GO}$ is the electronic position relative to the gauge origin position for the external magnetic potential.

The NR limits for the PV-NSR and PV-NMR-shielding tensors given in Eqs.~\eqref{eq:M-PV} and~\eqref{eq:sigma-PV}, respectively, were derived by some of us in Ref.~\citenum{AucarBorschevsky} by applying the linear response within the elimination of small components (LRESC) approach. In the same work, it was shown that the NR limits of the linear response functions involved in Eqs.~\eqref{eq:M-PV} and~\eqref{eq:sigma-PV} are exactly equal each other in the cases where $\bm{R}_{GO}$ is placed at the molecular CM. Following the LRESC model\cite{Agus2012,Agus_g_2014,Review-LRESC}, it is possible to expand these two relativistic second order properties in terms of the fine structure constant and to get their leading order relativistic corrections. Work in this line is currently in progress\cite{AucarPVLRESC}, and it can be shown that a close relationship between these properties also appears in the relativistic regime.

When the two enantiomers of a chiral molecule are analyzed, the isotropic nuclear spin-rotation constants $M_{X,iso}=\frac{1}{3}\textnormal{Tr}(\bm{M}_X)$ of their chiral centers $X$ will be given as the sum of a parity-conserving and a parity-violating contributions: $M_{X,iso}=M^{PC}_{X,iso} \pm M^{PV}_{X,iso}$, where the PV term of each of the two enantiomers have an opposite sign with respect to the other. This means that there will be a difference between the isotropic NSR constants of the same nuclei in the left(S)- and right(R)-handed enantiomers, which will be $|\Delta M_{X,iso}| = |M^S_{X,iso} - M^R_{X,iso}| = 2 \, |M^{PV}_{X,iso}|$. Similarly, $|g_X \, \Delta \sigma_{X,iso}| = |g_X \, \sigma^S_{X,iso} - g_X \, \sigma^R_{X,iso}| = 2 \, |g_X \, \sigma^{PV}_{X,iso}|$. In the following, we will report the $|\Delta M_{X,iso}|$ and the $\Delta \sigma_{X,iso}|$ quantities.

\section{Computational details}\label{sec:comp-det}

\begin{figure}
    \centering
    \includegraphics[scale=0.3]{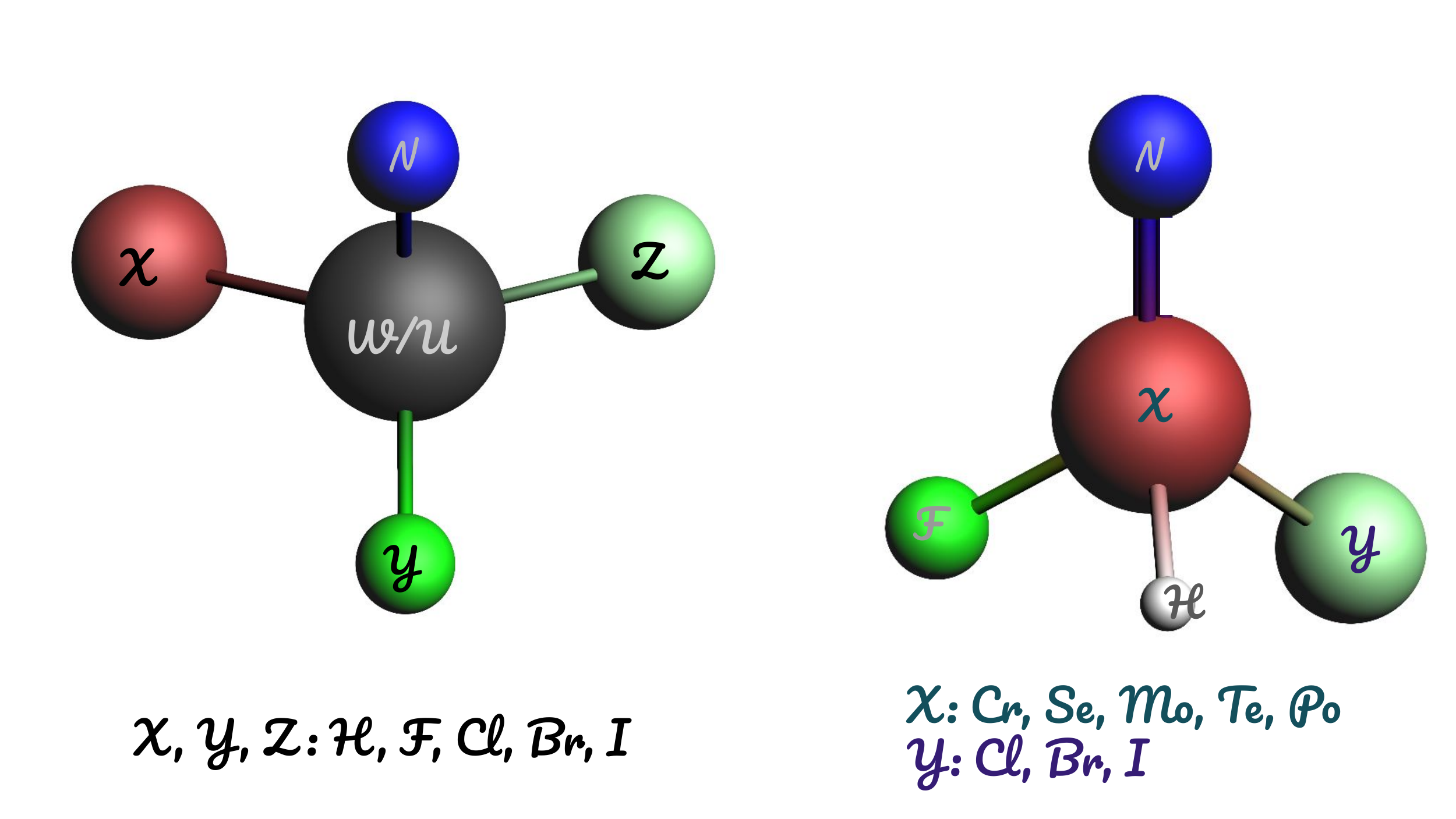}
    \caption{Schematic representation of the molecules studied in this work.}
    \label{fig:molecules}
\end{figure}

In this work we studied the PV-NSR and PV-NMR shielding tensors in the N$^{183}$W$XYZ$ and N$^{235}$U$XYZ$ series of molecules (with $X,Y,Z=$ $^1$H, $^{19}$F, $^{35}$Cl, $^{79}$Br, $^{127}$I), as well as in the N$X$HF$Z$ (with $X=$ $^{53}$Cr, $^{77}$Se, $^{95}$Mo, $^{125}$Te; $Z=$ $^{35}$Cl, $^{79}$Br, $^{127}$I) systems, schematically shown in Fig.~\ref{fig:molecules}. This selection of molecules allows us to analyze the effects produced by both the ligands and by the chiral centers on the studied properties.

We optimized all the molecular geometries using density functional theory (DFT), with the PBE0 functional\cite{PBE0}. Furthermore, due to the fact that specifically scalar relativistic effects are crucial for obtaining reliable molecular geometries\cite{pyykko1988relativistic,autschbach2012perspective,pyykko2008chemical}, we used scalar relativistic pseudo-potentials for the heavy atoms.

In the optimizations, Dunning's correlation consistent basis sets (aug-cc-pV5Z) were used on the light elements H, N, F\cite{dunning1989a,kendall1992a}, Cl\cite{woon1993a}, and Cr\cite{balabanov2005a}, and the Stuttgart-Koeln pseudo-potentials and correlation consistent basis sets (aug-cc-pV5Z-PP) on the heavy atoms, Br\cite{peterson2003b}, Mo\cite{peterson2007a}, I\cite{peterson2003b,peterson2006a}, and W\cite{figgen2009a}. These correlation consistent basis sets are not available for U, and thus we used the atomic natural orbital valence basis set in conjunction with the Stuttgart pseudo-potentials\cite{cao2003a,cao2004a,kuchle1994a}. The obtained structural parameters are shown in the Supplementary Material.

We performed all the optimizations using the default settings in the energy minimization scheme of the Gaussian program package\cite{Gaussian16}. Additionally, we performed a frequency analysis on the obtained geometries to confirm these as true minima. We computed the frequencies corresponding to the final geometries with the same method employed for the optimization and using the Gaussian code.

Regarding the parity-violation contributions to the properties analyzed in this work, to calculate $\bm{\sigma}^{PV}$ we used the \textsc{Dirac} program package\cite{DIRAC22,dirac-paper}, whereas a locally modified version of the same code was used to obtain the values of $\bm{M}^{PV}$. We used the DC and the LL Hamiltonians to obtain the relativistic and the NR results, respectively\cite{Saue2005}. The standard procedure to avoid the explicit calculation of (SS$\mid$SS) integrals in both the self-consistent field and the linear response blocks was followed in all the DC calculations, replacing these integrals by an energy correction. This choice is the default one in the \textsc{Dirac} code\cite{Visscher1997-SSSS}.

Unless otherwise stated, we have employed the Dyall's relativistic cv3z uncontracted basis sets (dyall.cv3z) for all the elements analyzed in this work\cite{KD02-3z, KD04, KD06, KD07-1, KD07-2, KD10-1, KD10-2, KD16}. An analysis of the basis set convergence of $M^{PV}_{iso}$ was performed for the tungsten nucleus in the NWHFI molecule and in Table \ref{tab:basis-conv-SR-dyall} we show the values of $M^{PV}_{W,iso}$ obtained using double-, triple-, and quadruple-zeta valence, core-valence, and all-electron (i.e., including correlating functions for all shells) uncontracted Dyall's basis sets (dyall.v$Y$z, dyall.cv$Y$z, and dyall.ae$Y$z, with $Y=$ 2, 3, 4)\cite{KD98,KD02-3z,KD04,KD06,KD10-1,KD12,KD16}. The value obtained using the dyall.cv3z basis set shows a good convergence and therefore this is the basis set we use across this work.

\begin{table}[htp]
\caption{\label{tab:basis-conv-SR-dyall} $M^{PV}_{iso}$ (in $\mu$Hz) for the $^{183}$W nucleus in the NWHFI molecule for different uncontracted Dyall's basis sets, using the DC Hamiltonian and the DFT/PBE0 approach.}
\centering
\begin{tabular*}{\linewidth}{@{\extracolsep{\fill}} l *{3}{c}}
\toprule
$X$ & dyall.v$X$z & dyall.cv$X$z & dyall.ae$X$z  \\
\midrule
2 & 33.29 & 33.40 & 33.40 \\ [0.5ex]
3 & 34.80 & \textbf{34.88} & 34.89 \\ [0.5ex]
4 & 35.02 & 35.07 & 35.08 \\
\bottomrule
\end{tabular*}
\end{table}

The common-gauge-origin approach was used in all the calculations, and the gauge origin for the external magnetic potential has been placed at the molecular center of mass. The small component basis sets were generated from the large component basis sets in all cases by applying the unrestricted kinetic balance (UKB) prescription\cite{GAA_JCP99}.

In the self-consistent-field calculations and for describing the NSD-PV perturbed Hamiltonian, a spherically symmetric Gaussian-type nuclear charge density distribution model was used\cite{Visscher1997}, as it has been shown that use of finite nuclear model is important to adequately describe the parity-conserving nuclear spin-rotation tensors\cite{Agus_NChDE_RSC2018}. However, a point-like nuclear magnetic moment $\bm{\mu}_N$ was employed in the term arising from the vector potential $\bm{A}_N= \frac{\mu_0}{4\pi} \bm{\mu}_N \times \frac{\bm{r}_N}{|\bm{r}_N|^3},$ where $\mu_0$ is the vacuum permeability, and $\bm{r}_N=\bm{r}-\bm{R}_N$ is the electronic position relative to nucleus $N$.

The response calculations were carried out at the 4c polarization propagator RPA level of theory employing the DHF and DKS-DFT wave functions.
DFT calculations were performed to study the influence of electronic correlation effects, and they were based on both the 4c-DC and NR-LL Hamiltonians. We used the NR exchange-correlation hybrid PBE0 functional\cite{PBE0}, because of its good performance in the 4c calculations of parity-conserving NSR constants (in comparison with experimental values)\cite{Agus_NChDE_RSC2018,Aucar_CH3X}. The same functional was recently used to calculate the PV-NSR tensors in a different series of molecules\cite{AucarBorschevsky}.

A deeper analysis of the electronic correlation effects was performed for the NWHF$X$ and NUHF$X$ series of molecules (with $X =$ Cl, Br, I). For these, we calculated the PV-NSR tensors using the following DFT functionals: (i) The local density approximation (LDA) functional\cite{LDA,LDA2}, (ii) the generalized gradient approximation (GGAs) PBE functional\cite{PBE}, and (iii) the hybrid functionals PBE0\cite{PBE0} and CAM-B3LYP\cite{CAMB3LYP}.

The response of Eq.~\eqref{eq:resp-eq} was solved with respect to the property gradient associated with the following operators: (i) the total electronic orbital and spin angular momenta operators, to calculate $\bm{M}_N^{PV}$; and (ii) the external magnetic field, for $\bm{\sigma}_N^{PV}$.

The employed nuclear $g$-factors used to calculate $\bm{\sigma}_N^{PV}$ were taken from Ref.~\citenum{Raghavan89} and their values are displayed in Table~\ref{tab:g-fac}.

\begin{table}[htp]
\centering
\caption{\label{tab:g-fac} Nuclear $g$-factors used in the calculations of $\bm{\sigma}_X^{PV}$.}
\begin{tabular*}{0.5\linewidth}{@{\extracolsep{\fill}} l c}
\toprule
$X$ & $g_X$ \\
\midrule
$^{53}$Cr  & $-0.31636$ \\
$^{77}$Se  & $1.070084$ \\
$^{95}$Mo  & $-0.36568$ \\
$^{125}$Te & $-1.77701$ \\
$^{183}$W  & $0.235569$ \\
$^{209}$Po & $1.376$    \\
$^{235}$U  & $-0.1085714286$ \\
\bottomrule
\end{tabular*}
\end{table}

\section{Results and discussions}\label{sec:results}

We report 4c relativistic calculations of isotropic PV-NSR and PV-NMR shielding constants. These linear response properties were obtained by applying the RPA approach of the polarization propagator theory, combined with DHF and DKS wave functions based on the DC Hamiltonian. The influence of electron correlation effects (taken as the difference between linear response calculations using DKS and DHF wave functions) as well as of relativistic effects on the calculated PV-NSR constants is investigated. We then proceed to analyze the effects of the chiral center and the chemical environment. These topics are addressed separately in the Subsections below.

\subsection{Correlation and relativistic effects}\label{sec:corr-rel}

In this Section we investigate the effect of relativity and electron correlation in the calculations of $M_{U,iso}^{PV}$ and $\sigma_{U,iso}^{PV}$. To analyze both effects, we consider the differences between the 4c and the NR values, and between the DFT/PBE0-based and the DHF-based calculations, respectively.

\begin{table*}[htp]
\caption{\label{tab:methods} Isotropic shifts of PV-NSR and ``normalized'' PV-NMR-shielding constants ($2\,|M^{PV}_{U,iso}|$, in $\mu$Hz, and $2\,|g_U\,\sigma^{PV}_{U,iso}|$, in $\mu$ppm, respectively) for the $^{235}$U nucleus in the NUHF$X$ ($X=$ Cl, Br, I) set of molecules, using different methods.}
\centering
\begin{tabular*}{\linewidth}{@{\extracolsep{\fill}} cl *{6}{c}}
\toprule
\multirow{2}{*}{Hamiltonian} & \multirow{2}{*}{Method} & \multicolumn{3}{c}{$2\,|M^{PV}_{U,iso}|$} & \multicolumn{3}{c}{$2\,|g_U\,\sigma^{PV}_{U,iso}|$} \\
 \cmidrule(lr{0.1in}){3-5} \cmidrule(lr{0.1in}){6-8}
 &  & $X=$ Cl & $X=$ Br & $X=$ I & $X=$ Cl & $X=$ Br & $X=$ I  \\
\midrule
\multirow{2}{*}{LL (NR)}  & DHF-RPA & 28.4   & 35.0  & 41.8  & -- & -- & -- \\ [0.5ex]
                     & DFT/PBE0     & 23.4   & 21.6  & 26.4  & --  & -- & -- \\ [0.5ex]
\midrule
\multirow{5}{*}{DC (4c)}  & DHF-RPA & 604.0  & 652.0 & 542.0 & 67.4   & 80.2   & 65.7  \\ [0.5ex]
                     & DFT/LDA      & 200.0  & 240.0 & 220.0 & 0.8    & 0.7    & 1.8   \\ [0.5ex]
                     & DFT/PBE      & 196.6  & 238.0 & 214.0 & 0.7    & 0.0    & 0.4   \\ [0.5ex]
                     & DFT/CAMB3LYP & 99.8   & 117.4 & 119.2 & 4.4    & 4.9    & 5.8   \\ [0.5ex]
                     & DFT/PBE0     & 121.4  & 142.6 & 141.0 & 2.8    & 3.5    & 5.4 \\
\bottomrule
\end{tabular*}
\end{table*}

\begin{figure*}
\centering
\begin{subfigure}{.48\textwidth}
  \centering
  \includegraphics[width=\linewidth]{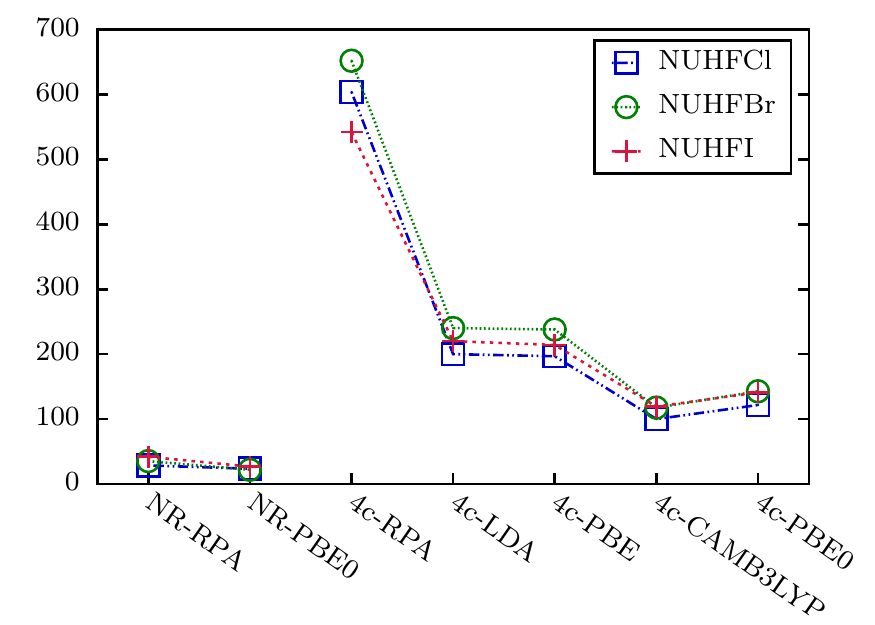}
  \caption{}
  \label{fig:SR-method}
\end{subfigure}
\begin{subfigure}{.48\textwidth}
  \centering
  \includegraphics[width=\linewidth]{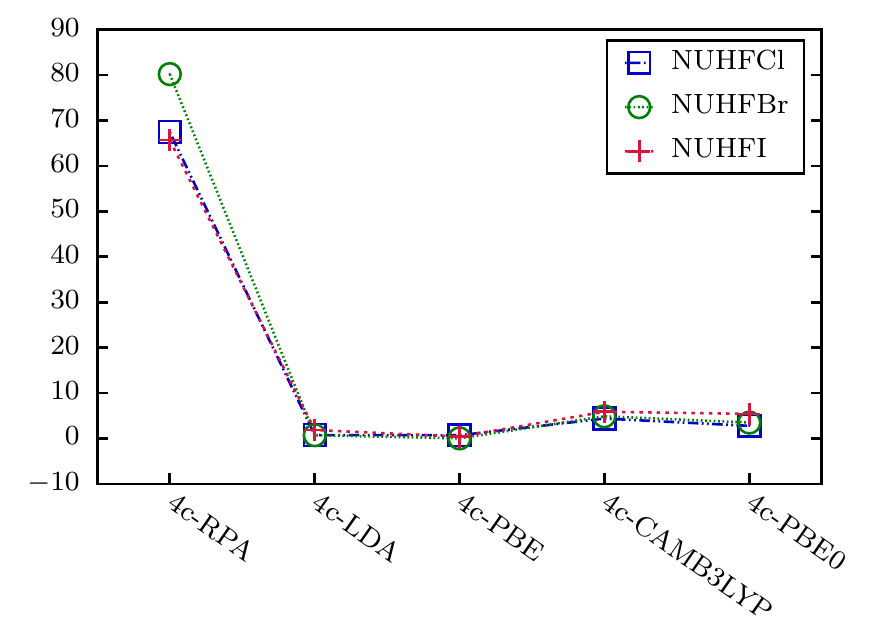}
  \caption{}
  \label{fig:sigma-method}
\end{subfigure}
\caption{Calculated values of (a) $|\Delta M_{U,iso}| = 2\,|M^{PV}_{U,iso}|$ (in $\mu$Hz) and (b) $|g_U \, \Delta \sigma_{U,iso}| = 2\,|g_U \sigma^{PV}_{U,iso}|$ (in $\mu$ppm) for $^{235}$U in NUHF$X$ molecular systems (with $X=$ Cl, Br, I) employing the LL and DC Hamiltonians, at the DHF, DFT-LDA, DFT-PBE, DFT-CAMB3LYP, and DFT-PBE0 levels of approach using the dyall.cv3z basis set for all elements.}
\label{fig:theory-method}
\end{figure*}

In Table~\ref{tab:methods} and Fig.~\ref{fig:SR-method}, the inclusion of relativity and electronic correlation are seen to have an opposite effect. At the DHF-RPA level of theory (4c-RPA vs NR-RPA), relativistic effects increase the values of $|\Delta M_{X,iso}^{PV}|$ significantly (with a 4c/NR ratio between 13 and 22). However, the calculated relativistic and NR values of $|\Delta M_{X,iso}^{PV}|$ are closer in magnitude when electron correlation is included using the DFT methodology (4c-PBE0 vs NR-PBE0), with a 4c/NR ratio of just between 5 and 7. The above observations demonstrate the importance of using a relativistic framework for meaningful investigations of these properties. Correlation effects are more pronounced in the relativistic than in the NR regime. For $ |g_U \, \Delta \sigma_{U,iso}^{PV}|$ (Fig.~\ref{fig:sigma-method}), the effect of electron correlation is also to decrease the calculated values significantly. Therefore, reliable calculations of isotropic PV-NSR and PV-NMR shielding constants require simultaneous inclusion of both relativistic and correlation effects.
Instabilities of the Kramer’s restricted DHF and DKS wave functions appear in the LL calculations of $\sigma_{U,iso}^{PV}$ and that is why they are not reported in this work. 

We also evaluate the dependence of the calculated properties on the chosen functional at the 4c-DFT level of theory. While $|g_U \, \Delta \sigma_{U,iso}^{PV}|$ is stable under the change of the DFT functional, $|\Delta M_{U,iso}^{PV}|$ varies more significantly (see Figs.~\ref{fig:SR-method} and \ref{fig:sigma-method}).

\begin{figure*}
\centering
\begin{subfigure}{.48\textwidth}
  \centering
  \includegraphics[width=\linewidth]{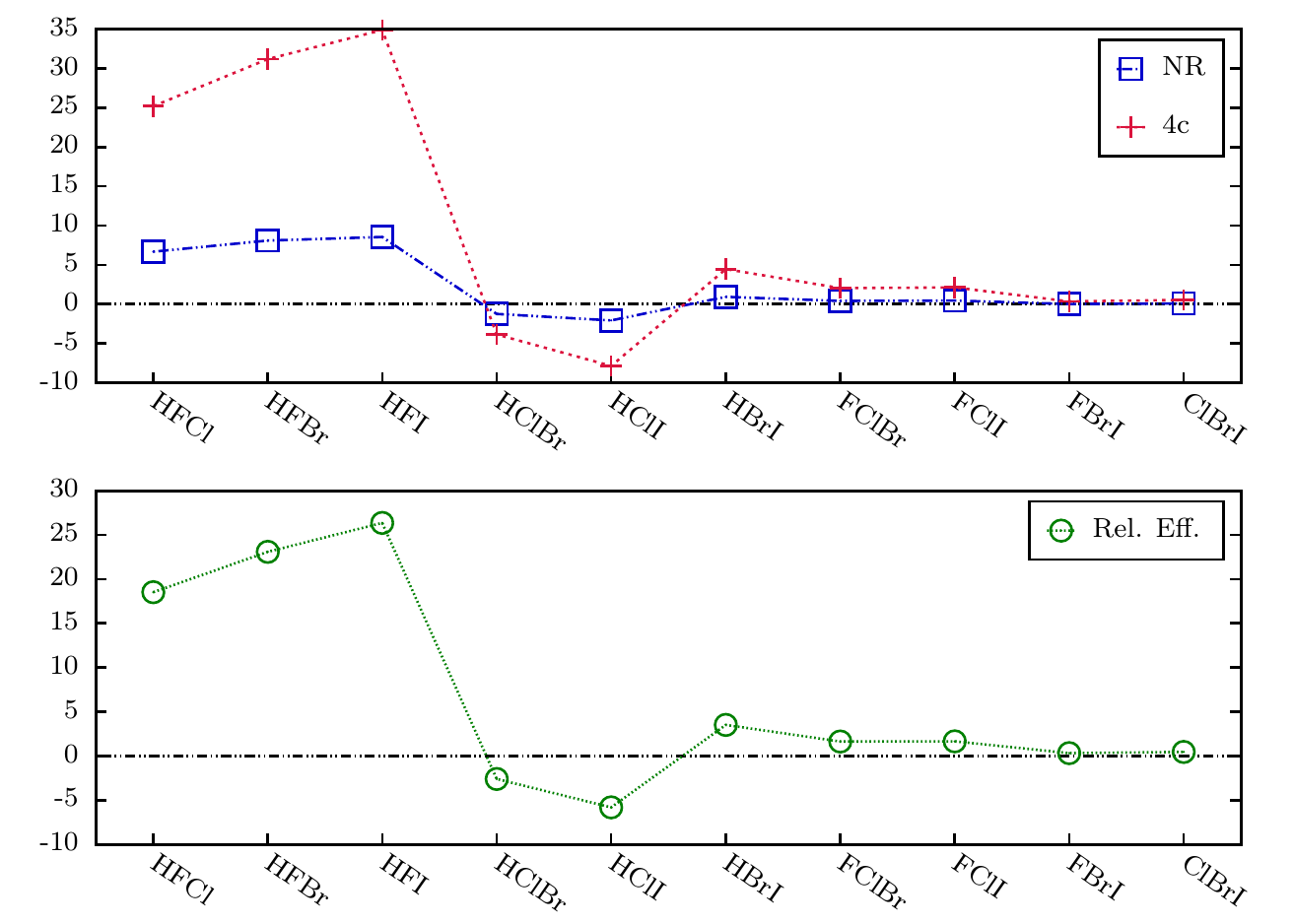}
  \caption{}
  \label{fig:rel-eff-W}
\end{subfigure}
\begin{subfigure}{.48\textwidth}
  \centering
  \includegraphics[width=\linewidth]{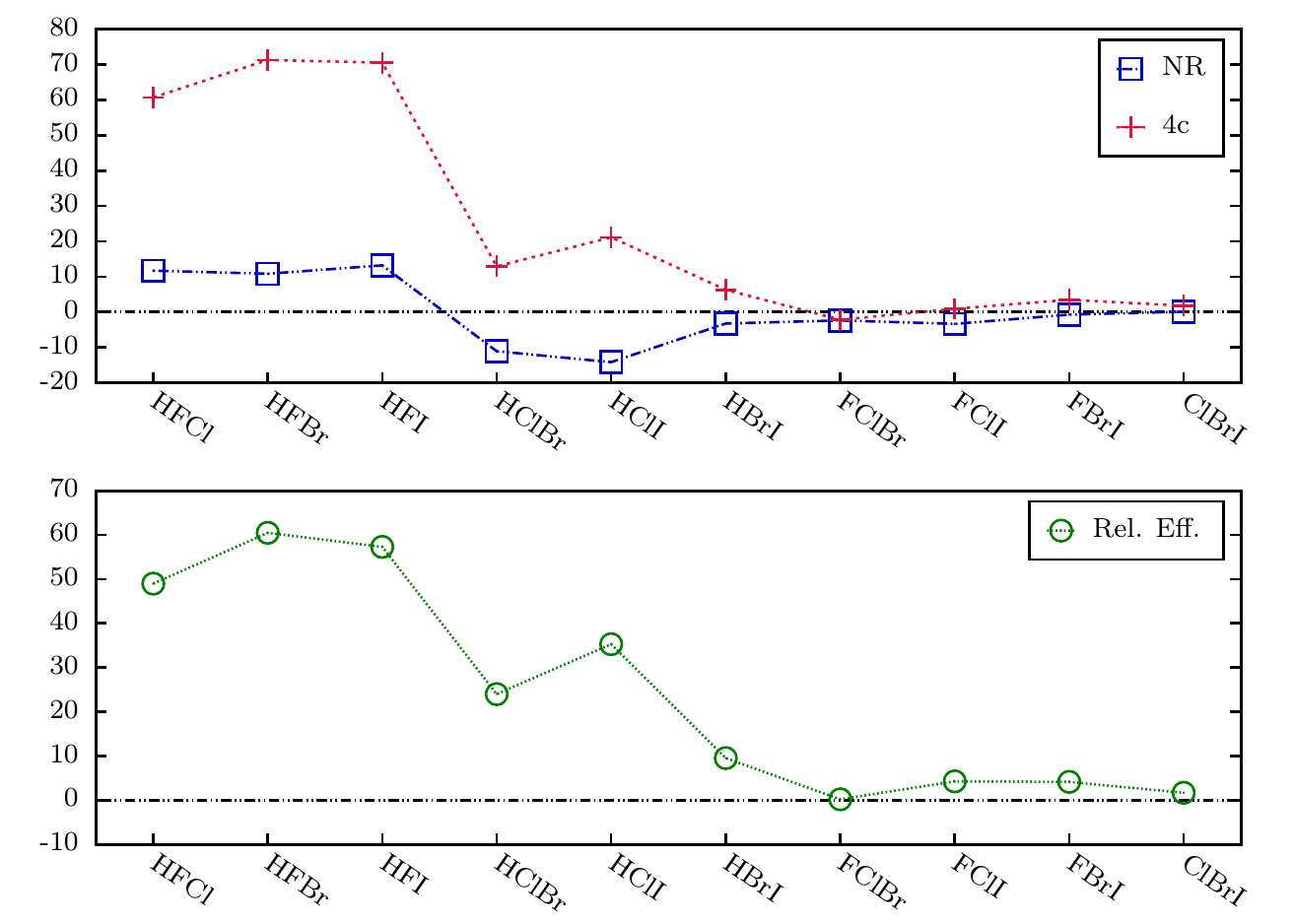}
  \caption{}
  \label{fig:rel-eff-U}
\end{subfigure}
\caption{(Top) Calculated values of $M^{PV}_{iso}$ for (a) $^{183}$W and (b) $^{235}$U in the NW$XYZ$ and NU$XYZ$ systems ($X,Y,Z=$ H, F, Cl, Br, I), respectively, employing the LL and DC Hamiltonians at the DFT-PBE0 level of theory and using the dyall.cv3z basis set for all elements. (Bottom) Relativistic effects taken as the difference between the 4c and NR values of $M^{PV}_{iso}$. All values are in $\mu$Hz.}
\end{figure*}

Figs.~\ref{fig:rel-eff-W} and~\ref{fig:rel-eff-U} show the relativistic effects at the DFT/PBE0 level of theory in the calculations of $M_{iso}^{PV}$ for the tungsten and uranium nuclei in the NW$XYZ$ and NU$XYZ$ systems (with $X,Y,Z=$ H, F, Cl, Br, I), respectively. The H- and F-containing systems exhibit the highest isotropic PV-NSR constants in both the NR and 4c regimes.

\begin{figure}
\centering
  \includegraphics[width=\linewidth]{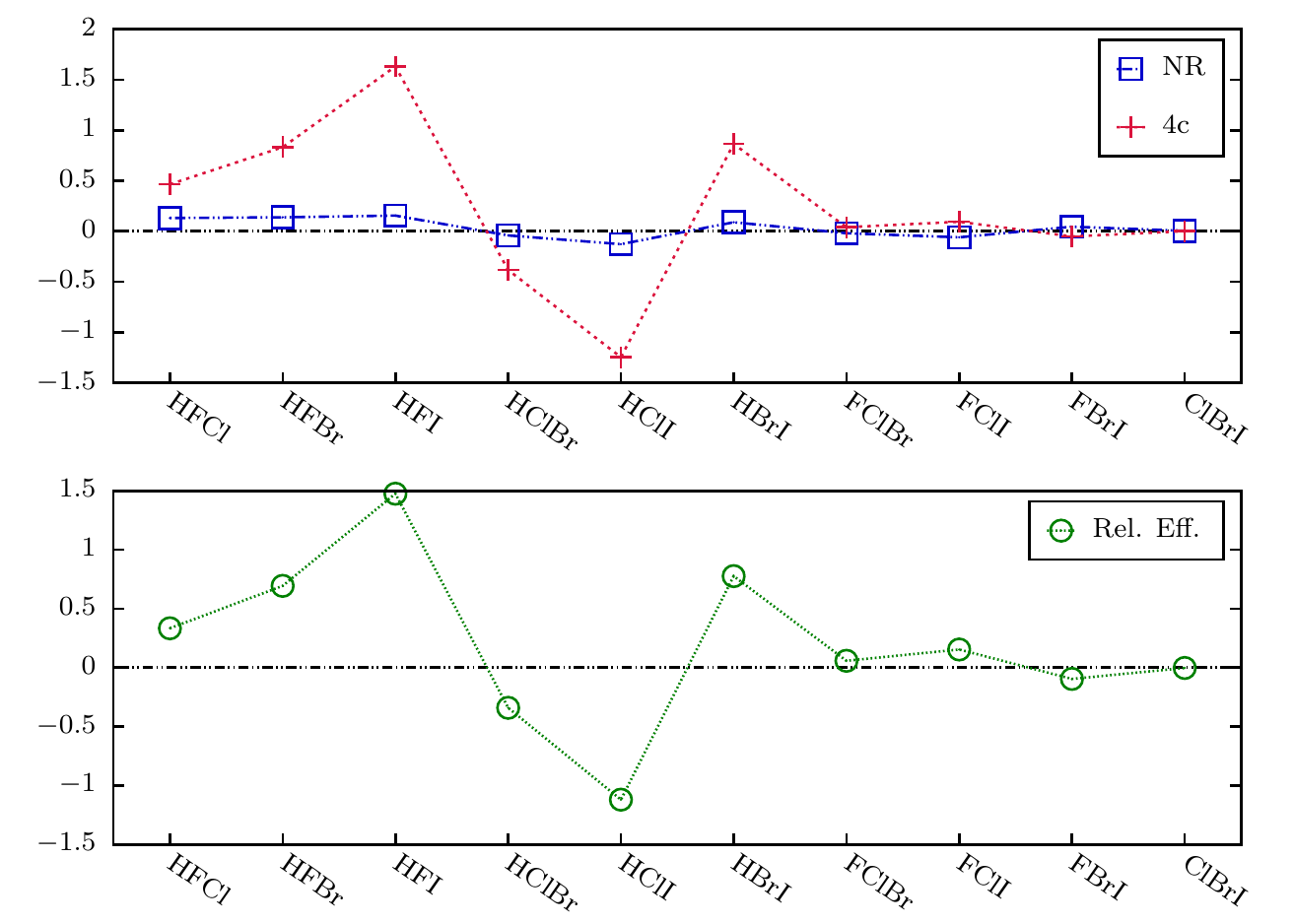}
\caption{(Top) Calculated values of $g_W \, \sigma^{PV}_{W,iso}$ for the $^{183}$W nucleus in the NW$XYZ$ systems ($X,Y,Z=$ H, F, Cl, Br, I), employing the LL and DC Hamiltonians at the DFT-PBE0 level of theory and using the dyall.cv3z basis set for all elements. (Bottom) Relativistic effects taken as the difference between the 4c and NR values of $g_W \, \sigma^{PV}_{W,iso}$. All values are given in $\mu$ppm.}
\label{fig:rel-eff-shi-W}
\end{figure}

In Fig.~\ref{fig:rel-eff-shi-W} a similar behavior can be observed for $g_W \, \sigma^{PV}_{W,iso}$ in the NW$XYZ$ set of molecules (with $X,Y,Z=$ H, F, Cl, Br, I).

\subsection{Fixing chiral centers. Analysis of the environment}\label{sec:fix-center}

Initially, we focus our analysis on the effects of the ligands on the size of the calculated $| \Delta M^{PV}_{K,iso}|$ and $|g_K \, \Delta \sigma^{PV}_{K,iso}|$ ($K=$ $^{183}$W and $^{235}$U) on the NW$XYZ$ and NU$XYZ$ series of molecules ($X,Y,Z=$ H, F, Cl, Br, I), as seen in Table~\ref{tab:ligands}.

\begin{table}[htp]
\caption{\label{tab:ligands} Calculations of $2\,|M^{PV}_{K,iso}|$ (in $\mu$Hz) and $2\,|g_K\,\sigma^{PV}_{K,iso}|$ (in $\mu$ppm) for the $K=$ $^{183}$W and $^{235}$U nuclei in the N$KXYZ$ ($X,Y,Z=$ H, F, Cl, Br, I) molecules. The DC Hamiltonian was used at the DFT/PBE0 level of approach.}
\centering
\begin{tabular*}{\linewidth}{@{\extracolsep{\fill}} l *{4}{c}}
\toprule
  & \multicolumn{2}{c}{$2\,|M^{PV}_{K,iso}|$ ($\mu$Hz)}& \multicolumn{2}{c}{$2\,|g_K\,\sigma^{PV}_{K,iso}|$ ($\mu$ppm)} \\
\cmidrule(lr{0.1in}){2-3} \cmidrule(lr{0.1in}){4-5}
$XYZ$   & $K=$ W & $K=$ U & $K=$ W & $K=$ U \\
\midrule
HFCl  & 50.40 & 121.40& 0.93  & 2.77  \\ [0.5ex]
HFBr  & 62.40 & 142.60& 1.66  & 3.48  \\ [0.5ex]
HFI   & 69.80 & 141.00& 3.26  & 5.40  \\ [1.5ex]
%
%
HClBr & 7.70  & 25.80 & 0.77  & 0.84  \\ [0.5ex]
HClI  & 15.74 & 42.20 & 2.50  & 3.66  \\ [0.5ex]
HBrI  & 0.90  & 12.56 & 1.73  & 2.54  \\ [1.5ex]
%
%
FClBr & 4.10  & 4.36  & 0.08  & 0.38  \\ [0.5ex]
FClI  & 4.24  & 1.79  & 0.19  & 0.81  \\ [0.5ex]
FBrI  & 0.67  & 6.90  & 0.10  & 0.31  \\ [0.5ex]
ClBrI & 1.08  & 3.52  & 0.00  & 0.06  \\
\bottomrule
\end{tabular*}
\end{table}

\begin{figure*}
\centering
\begin{subfigure}{.48\textwidth}
  \centering
  \includegraphics[width=\linewidth]{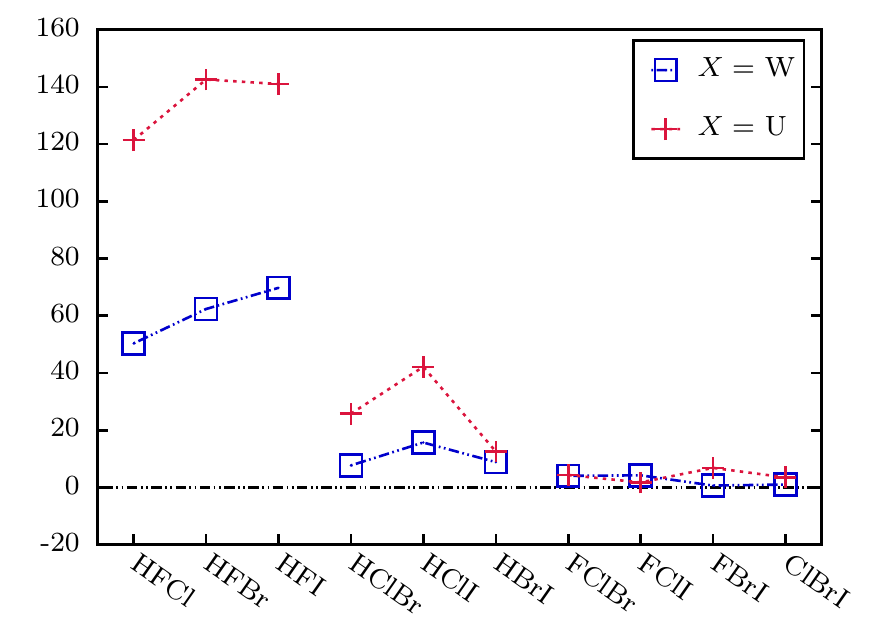}
  \caption{}
  \label{fig:SR-ligands}
\end{subfigure}
\begin{subfigure}{.48\textwidth}
  \centering
  \includegraphics[width=\linewidth]{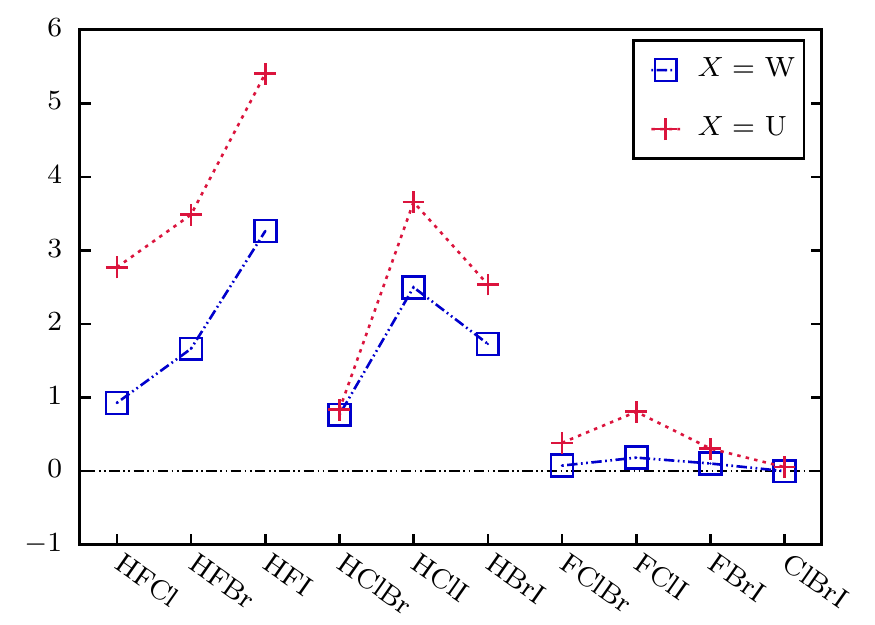}
  \caption{}
  \label{fig:sigma-ligands}
\end{subfigure}
\caption{Calculated values of (a) $|\Delta M_{X,iso}| = |M^S_{X,iso} - M^R_{X,iso}| = 2\,|M^{PV}_{X,iso}|$ (in $\mu$Hz) 
(b) $|g_X \, \Delta \sigma_{X,iso}| = 
2\,|g_X\,\sigma^{PV}_{X,iso}|$ (in $\mu$ppm) for the $X$ nuclei in N$XABC$ systems (with $X=$ $^{183}$W, $^{235}$U; and $A,B,C=$ H, F, Cl, Br, I), employing the DC Hamiltonian at the DFT-PBE0 level of theory and using the dyall.cv3z basis set for all the elements.}
\label{fig:fix-center}
\end{figure*}

In Fig.~\ref{fig:fix-center} we have classified all the systems in three different groups: systems containing (i) both H and F atoms, (ii) only H atoms without fluorine, and (iii) systems containing no H atoms. In Fig.~\ref{fig:SR-ligands} it is seen that the size of $|\Delta M^{PV}_{X,iso}|$ follows (i) $>$ (ii) $>$ (iii).

As stated in Sec.~\ref{sec:corr-rel}, the H- and F-containing systems exhibit the highest isotropic PV-NSR constants. This can be due to the fact that an increasingly asymmetric electronic distribution appears in the molecular chiral center when the difference of electronegativities between two of the ligands also increases. The electronegativities of the atoms considered in this work follow the tendency $\chi_\text{F}>\chi_\text{Cl}>\chi_\text{Br}>\chi_\text{I}>\chi_\text{H}$. It implies that F and H have the largest and smallest electronegativities, respectively, and hence the largest PV effects are found for the NWHF$X$ and NUHF$X$ molecules (with $X=$ Cl, Br, I). On the other hand, as Cl and Br have similar electronegativities, the presence of these ligands will not generate a strong asymmetry in the electronic cloud at the chiral center position, and so the resulting PV effects will be between the lowest for the studied series of molecules, in spite of them being heavier than the HF-containing molecules.

On the other hand, $\sigma^{PV}_{X,iso}$ does not follow the same tendency, as seen in Fig.~\ref{fig:sigma-ligands}. For $|g_X \, \Delta \sigma^{PV}_{X,iso}|$, groups (i) and (ii) show a more similar behavior, while the systems that do not contain hydrogen atoms (i.e., those of group (iii)) still yield the lowest values.
The effect of the environment is remarkably large for these properties, and more significant than the atomic number of the chiral center, in many cases. For example, $|\Delta M^{PV}_{X,iso}|$ is much larger in NWHFI than in the heavier NUClBrI. Therefore, we can identify the  H- and F-containing systems as the most promising candidates for measurements in the investigated series. Similar conclusions were reached in an earlier paper that investigated the shieldings in the NW$XYZ$ compounds\cite{Nahrwold2014}.

\subsection{Fixing environment. Analysis of chiral centers}

In general, U-containing systems have higher PV-NSR and PV-NMR shielding constants than W-containing systems (for the same ligand environment). Furthermore, in both cases there is not a significant effect of the ligand when it is different to H and F.
Therefore, we proceeded to study the effect of the chiral center while keeping the H and F ligands in the systems under study.

To analyze the effects produced by the PV interactions as a function of the chiral center, we included in the study further molecules containing elements with 6 valence electrons, and classified our sample into two groups according to their metal valence open-shell orbitals: $p$-group elements: Se, Te, Po ($s^2p^4$) and transition metals: Cr, Mo ($d^5s^1$), W ($d^4s^2$), and U ($f^3d^1s^2$). The obtained results are displayed in Table~\ref{tab:centers}.

\begin{table}[htp]
\caption{\label{tab:centers} Calculations of $2\,|M^{PV}_{K,iso}|$ (in $\mu$Hz) and $2\,|g_K\,\sigma^{PV}_{K,iso}|$ (in $\mu$ppm) for the $K$ nuclei (with $K=$ $^{53}$Cr, $^{77}$Se, $^{95}$Mo, $^{125}$Te, $^{183}$W, $^{209}$Po, and $^{235}$U) in the N$K$HF$X$ ($X$= Cl, Br, I) series of molecules. The DC Hamiltonian was used in the calculations at the DFT/PBE0 level of approach. DFT/PBE was used the for Po-containing molecules to avoid instabilities of the Kramer’s restricted DKS wave functions appearing in the DFT/PBE0 calculations of these systems.}
\centering
\begin{threeparttable}
\begin{tabular*}{\linewidth}{@{\extracolsep{\fill}} l *{6}{c}}
\toprule
   & \multicolumn{3}{c}{$2\,|M^{PV}_{K,iso}|$ ($\mu$Hz)} & \multicolumn{3}{c}{$2\,|g_K\,\sigma^{PV}_{K,iso}|$ ($\mu$ppm)} \\
\cmidrule(lr{0.1in}){2-4} \cmidrule(lr{0.1in}){5-7}
$K$   & N$K$HFCl  & N$K$HFBr & N$K$HFI & N$K$HFCl  & N$K$HFBr & N$K$HFI \\
\midrule
Cr & 0.52 & 0.61 & 0.59 & 0.02 & 0.02 & 0.10 \\ [0.5ex]
Se & 6.34 & 9.76 & 10.40 & 0.08 & 0.08 & 0.04 \\ [0.5ex]
Mo & 6.18 & 7.96 & 8.86 & 0.09 & 0.20 & 0.47 \\ [0.5ex]
Te & 14.72 & 25.80 & 28.20 & 1.07 & 1.41 & 1.26 \\ [0.5ex]
W & 50.40 & 62.40 & 69.80 & 0.93 & 1.66 & 3.26 \\ [0.5ex]
Po & 286.00\tnotex{pbe} & 478.00\tnotex{pbe} & 600.00\tnotex{pbe} & 3.94\tnotex{pbe} & 3.26\tnotex{pbe} & 10.34\tnotex{pbe} \\ [0.5ex]
U & 121.40 & 142.60 & 141.00 & 2.77 & 3.48 & 5.40 \\
\bottomrule
\end{tabular*}
\begin{tablenotes}
\item[a] \label{pbe} Calculated using the DFT/PBE functional.
\end{tablenotes}
\end{threeparttable}
\end{table}

\begin{figure*}
\centering
\begin{subfigure}{.32\textwidth}
  \centering
  \includegraphics[width=\linewidth]{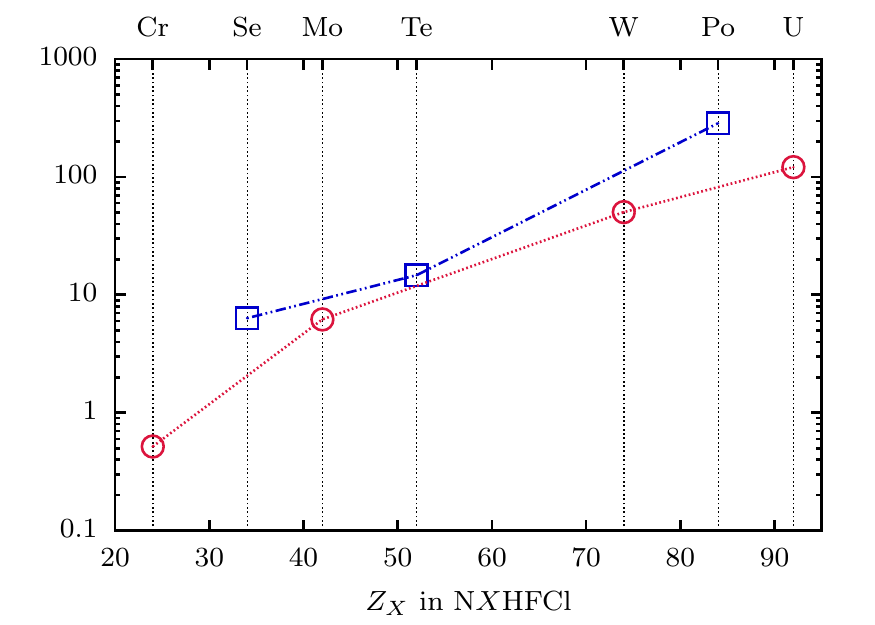}
  \caption{}
  \label{fig:SR-fix-ligands-Cl}
\end{subfigure}
\begin{subfigure}{.32\textwidth}
  \centering
  \includegraphics[width=\linewidth]{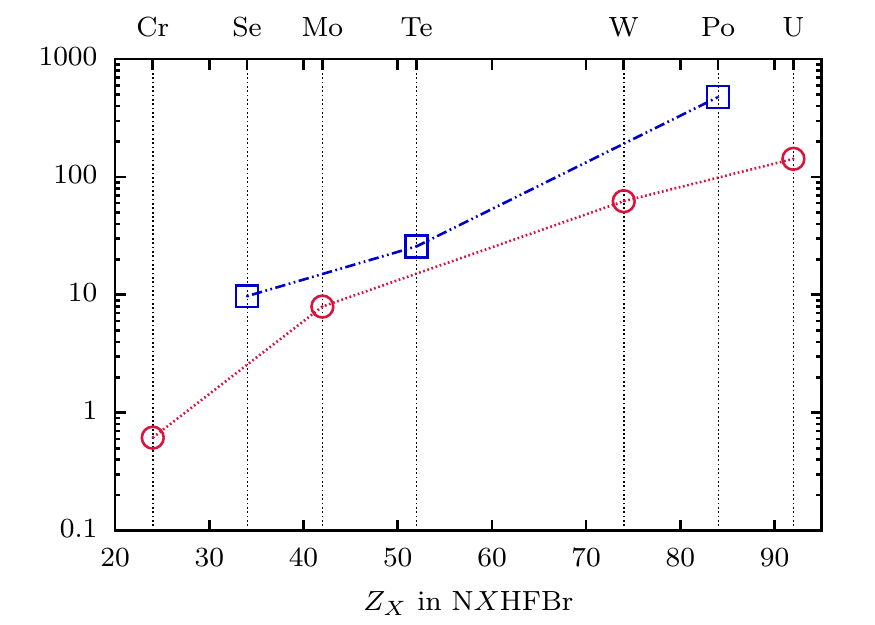}
  \caption{}
  \label{fig:SR-fix-ligands-Br}
\end{subfigure}
\begin{subfigure}{.32\textwidth}
  \centering
  \includegraphics[width=\linewidth]{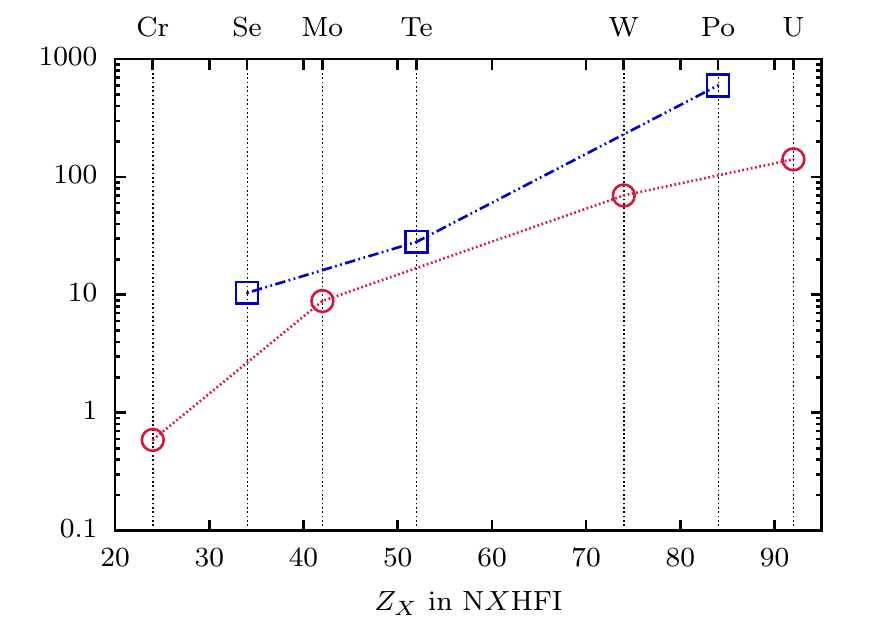}
  \caption{}
  \label{fig:SR-fix-ligands-I}
\end{subfigure}
\caption{Calculated values of $|\Delta M_{X,iso}| = 2\,|M^{PV}_{X,iso}|$ (in $\mu$Hz) for the $X$ nuclei in N$X$HF$Y$ systems with $X=$ $^{53}$Cr, $^{77}$Se, $^{95}$Mo, $^{125}$Te, $^{183}$W, $^{209}$Po, and $^{235}$U, for $Y=$ (a) Cl, (b) Br, and (c) I, employing the DC Hamiltonian at the DFT-PBE0 level of theory (except for NPoHF$X$, where the DFT-PBE functional was used instead) and using the dyall.cv3z basis set for all the elements.}
\label{fig:SR-fix-ligands}
\end{figure*}

\begin{figure*}
\centering
\begin{subfigure}{.32\textwidth}
  \centering
  \includegraphics[width=\linewidth]{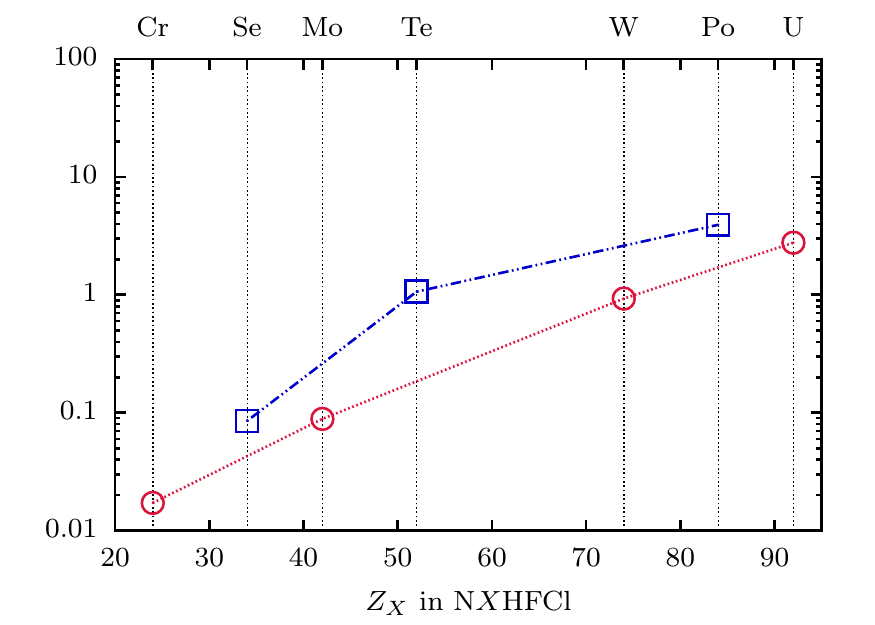}
  \caption{}
  \label{fig:shi-fix-ligands-Cl}
\end{subfigure}
\begin{subfigure}{.32\textwidth}
  \centering
  \includegraphics[width=\linewidth]{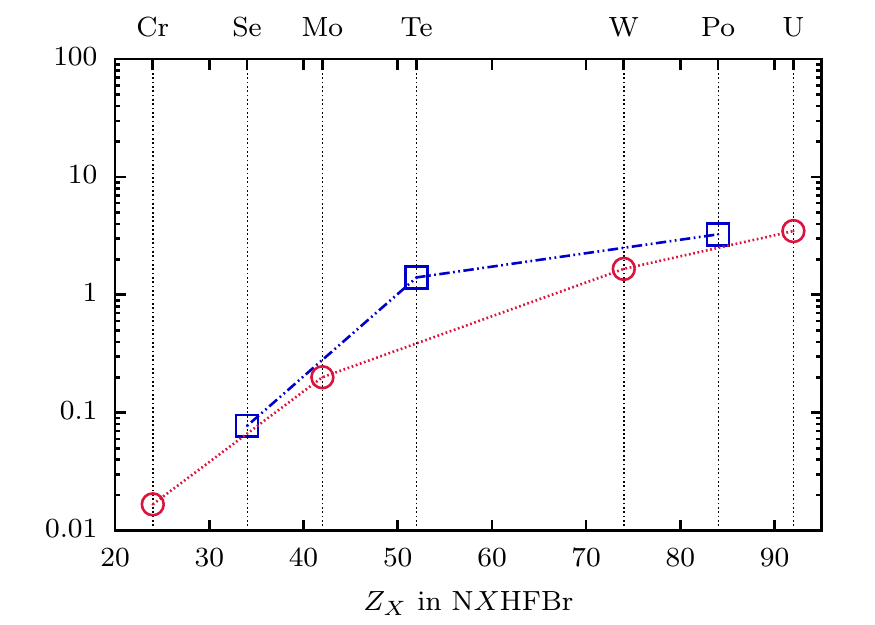}
  \caption{}
  \label{fig:shi-fix-ligands-Br}
\end{subfigure}
\begin{subfigure}{.32\textwidth}
  \centering
  \includegraphics[width=\linewidth]{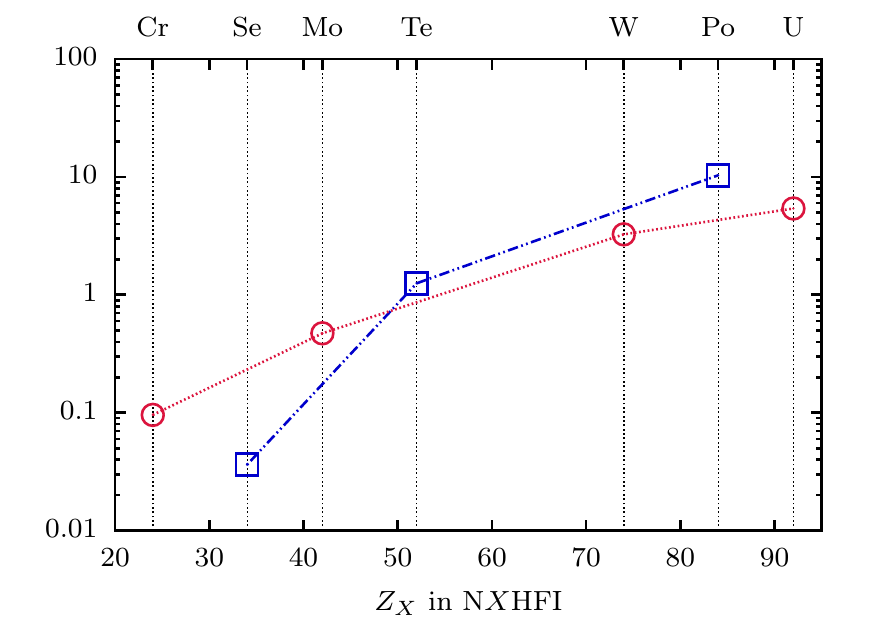}
  \caption{}
  \label{fig:shi-fix-ligands-I}
\end{subfigure}
\caption{Calculated values of $|g_X \, \Delta \sigma_{X,iso}| = 2\,|g_X \, \sigma^{PV}_{X,iso}|$ (in $\mu$ppm) for the $X$ nuclei in N$X$HF$Y$ systems with $X=$ $^{53}$Cr, $^{77}$Se, $^{95}$Mo, $^{125}$Te, $^{183}$W, $^{209}$Po, and $^{235}$U, for $Y=$ (a) Cl, (b) Br, and (c) I, employing the DC Hamiltonian at the DFT-PBE0 level of theory (except for NPoHF$X$, where the DFT-PBE functional was used instead) and using the dyall.cv3z basis set for all the elements.}
\label{fig:shi-fix-ligands}
\end{figure*}

Figs.~\ref{fig:SR-fix-ligands} and~\ref{fig:shi-fix-ligands} present the values of $|\Delta M_{X,iso}|$ and $|g_X\ \Delta \sigma_{X,iso}|$ for the $X$ nuclei in N$X$HF$Y$ systems with $X=$ $^{53}$Cr, $^{77}$Se, $^{95}$Mo, $^{125}$Te, $^{183}$W, $^{209}$Po, and $^{235}$U, and $Y=$ Cl, Br, I as a function of the atomic number of the chiral center, respectively. As expected, the PV-NSR and PV-NMR-shielding constants increase with the atomic number; however, it is interesting to remark that this dependency is different for the two groups of molecules. Generally, the PV-NSR and PV-NMR-shieldings increase faster with the atomic number in the systems containing $p$ open shell orbitals than in those with $d$ open shells.
We found that  the  excitations from  the valence shells provide the highest contributions to these properties in the linear responses of Eqs.~\eqref{eq:M-PV} and \eqref{eq:sigma-PV}. We can thus explain the difference in the trends by the fact that the valence electrons in the $p$ orbitals experience higher relativistic effects and contract more than those in $d$ and $f$ shells, bringing the electrons closer to the nucleus
This leads us to find the largest PV effects in Po-containing systems. Notice that we have included systems containing U (heavier than Po) in our study.

On the other hand, Figs.~\ref{fig:SR-fix-ligands-Cl}, \ref{fig:SR-fix-ligands-Br}, \ref{fig:SR-fix-ligands-I}, and also Figs.~\ref{fig:shi-fix-ligands-Cl}, \ref{fig:shi-fix-ligands-Br}, and \ref{fig:shi-fix-ligands-I} show the same dependency of both properties on the metal atomic number when $Y$ is varied in N$X$HF$Y$ molecules. There is only one exception to this tendency, for $|g_{\text{Se}}\ \Delta \sigma_{\text{Se},iso}|$ in NSeHFI.

\section{Conclusions}\label{sec:conclusion}

NSD-PV contributions to the NSR and NMR shielding tensors were obtained in this work using the relativistic expression proposed in Ref.~\citenum{AucarBorschevsky} for a series of light- and heavy-element-containing tetrahedral molecules. We have shown that relativity plays a crucial role in describing the PV effects in these parameters for the chiral centers of the N$AXYZ$ series of molecules (with $A=$ Cr, Mo, W, Se, Te, Po, U; and $X,Y,Z=$ H, F, Cl, Br, I). Electron correlation effects are as important as relativity for describing the PV-NSR and PV-NMR-shielding constants in these molecules, and are more pronounced in the relativistic than in the NR calculations. Therefore, reliable calculations of $M_{iso}^{PV}$ and $\sigma_{iso}^{PV}$ require simultaneous inclusion of both relativistic and correlation effects.

We have shown that molecules containing both H and F atoms exhibit the highest isotropic PV-NSR and PV-NMR shielding constants, making this choice of ligand the most promising for such investigations.

We also provide a study of the effect of the chiral center (keeping the H and F ligands). Here, we included further tetrahedral molecules containing elements with 6 valence electrons as chiral centers. The isotropic PV-NSR and PV-NMR-shielding constants increase with the atomic number of the chiral center. However, depending on the electronic structure of the metal atom, different tendencies emerge; that is the PV-NSR and PV-NMR-shieldings increase faster with the atomic number in systems containing $p$ open-shell orbitals than in those with $d$ open shells. We  found the highest value of PV-NSR constant shift for the $^{209}$Po nucleus in NPoHFI, on the order of 0.6~mHz.

While the PV-NSR constants given in this work do not reach the experimental sensitivity limit for NSR constants, on the order of 1 to 10~Hz\cite{Giuliano2011,Yoo2016}, they give crucial insights for the experimental search of PV effects in molecules. A natural extension of this work is to search for promising realistic molecules for measurements that are both experimentally accessible and benefit from larger PV-NSR contributions. We will use the insights from this and earlier works on PV-NMR-shieldings\cite{Berger2007,Zanasi2007,Figgen2008,Weijo2008,Weijo2009,Nahrwold2014,Eills2017} as a starting point to identify new molecular candidates for further computational investigations.

\section*{\label{sec:sup-mat}Supplementary Material}
See supplementary material for the optimized structural parameters of the chiral molecules studied in this work.

\begin{acknowledgments}
This work has been performed under the Project HPC-EUROPA3 (INFRAIA-2016-1-730897), with the support of the EC Research Innovation Action under the H2020 Programme; in particular, IAA gratefully acknowledges the support of the University of Groningen and the computer resources and technical support provided by SURFsara.
We would like to thank as well the Center for Information Technology of the University of Groningen in The Netherlands, and the Institute for Modeling and Innovation on Technologies (IMIT) of Argentina for their support and for providing access to the Peregrine and IMIT high performance computing clusters. This work also made use of the Dutch national e-infrastructure with the support of the SURF Cooperative using grant no.~EINF-3247.
IAA acknowledges partial support from FONCYT by grants PICT-2016-2936 and PICT-2020-SerieA-00052. This publication is part of the project \textit{High Sector Fock space coupled cluster method: benchmark
accuracy across the periodic table}
(with project number VI.Vidi.192.088 of the research programme Vidi which is  financed by the Dutch Research Council (NWO)).
\end{acknowledgments}

\bibliography{PV-SR-NAXYZ}

\end{document}